\documentclass[aps,prl,twocolumn,showpacs,showkeys,footinbib,amsmath,amssymb,superscriptaddress]{revtex4-2}
\usepackage{amsmath}
\usepackage{amssymb}
\usepackage{graphicx}
\usepackage{bm}
\usepackage{color}
\usepackage{relsize}
\usepackage{braket}
\usepackage{bbold}
\usepackage{txfonts}
\usepackage{epstopdf}
\usepackage{soul}
\usepackage{color,xcolor}
\usepackage{float}
\PassOptionsToPackage{colorlinks=true,
	linkcolor=blue,
	citecolor=blue,
	urlcolor=blue}{hyperref}
\usepackage{hyperref}

\begin{document}

\title{Large Field-Free Superconducting Diode Effect with Nonmonotonic Polarity Reversals in NbSe$_2$/CrBr$_3$ Heterostructures}

\author{Shengbiao Sun}
\thanks{These authors contributed equally to this work.}
\affiliation{Southern University of Science and Technology, Shenzhen, 518055, China}
\affiliation{International Quantum Academy, and Shenzhen Branch, Hefei National Laboratory, Shenzhen, 518048, China}

\author{Lang Xiao}
\thanks{These authors contributed equally to this work.}
\affiliation{Southern University of Science and Technology, Shenzhen, 518055, China}
\affiliation{International Quantum Academy, and Shenzhen Branch, Hefei National Laboratory, Shenzhen, 518048, China}

\author{Chenghao Shen}
\thanks{These authors contributed equally to this work.}
\affiliation{Eastern Institute for Advanced Study, Eastern Institute of Technology, Ningbo, Zhejiang 315200, China}

\author{Bowen Hao}
\thanks{These authors contributed equally to this work.}
\affiliation{Eastern Institute for Advanced Study, Eastern Institute of Technology, Ningbo, Zhejiang 315200, China}

\author{Jia‑Peng Peng}
\affiliation{Southern University of Science and Technology, Shenzhen, 518055, China}
\affiliation{International Quantum Academy, and Shenzhen Branch, Hefei National Laboratory, Shenzhen, 518048, China}

\author{Ziye Zhu}
\affiliation{Eastern Institute for Advanced Study, Eastern Institute of Technology, Ningbo, Zhejiang 315200, China}

\author{Feiyue Wang}
\affiliation{State Key Laboratory of Optoelectronic Materials and Technologies, Sun Yat‑sen University, Guangzhou, 510275, China}

\author{Qilin Han}
\affiliation{International Quantum Academy, and Shenzhen Branch, Hefei National Laboratory, Shenzhen, 518048, China}
\affiliation{State Key Laboratory of Optoelectronic Materials and Technologies, Sun Yat‑sen University, Guangzhou, 510275, China}

\author{Ya‑Qing Bie}
\affiliation{State Key Laboratory of Optoelectronic Materials and Technologies, Sun Yat‑sen University, Guangzhou, 510275, China}

\author{Shuo Wang}
\email{wangshuo@iqasz.cn}
\affiliation{International Quantum Academy, and Shenzhen Branch, Hefei National Laboratory, Shenzhen, 518048, China}

\author{Tong Zhou}
\email{tzhou@eitech.edu.cn}
\affiliation{Eastern Institute for Advanced Study, Eastern Institute of Technology, Ningbo, Zhejiang 315200, China}

\author{Dapeng Yu}
\affiliation{International Quantum Academy, and Shenzhen Branch, Hefei National Laboratory, Shenzhen, 518048, China}

\author{Ben‑Chuan Lin}
\email{linbenchuan@iqasz.cn}
\affiliation{International Quantum Academy, and Shenzhen Branch, Hefei National Laboratory, Shenzhen, 518048, China}

\begin{abstract}
The superconducting diode effect (SDE), characterized by nonreciprocal dissipationless supercurrent, offers a promising route toward ultralow-power superconducting electronics. Yet most realizations require an external magnetic field, and achieving a large, controllable SDE at zero field remains challenging. Here we report a field-free SDE with an efficiency reaching 35.7\% in a van der Waals NbSe$_2$/CrBr$_3$ heterostructure, whose polarity is programmable by magnetic-field history. Unlike conventional mechanisms based on finite-momentum pairing, the SDE originates from the leading symmetry-allowed cubic term odd in Cooper-pair momentum, which yields unequal critical currents while leaving the equilibrium condensate at zero momentum. Remarkably, upon sweeping a perpendicular magnetic field, the diode polarity undergoes nonmonotonic and hysteretic reversals that cannot be explained by Meissner screening or conventional ferromagnetic proximity. Combining transport measurements, micromagnetic simulations, and a generalized Ginzburg–Landau theory, we attribute these unconventional behaviors to layered ferrimagnetism in CrBr$_3$, where coexisting ferromagnetic and antiferromagnetic interlayer couplings produce a history-dependent interfacial exchange field acting on NbSe$_2$. Our results reveal a distinct mechanism for nonreciprocal superconductivity and establish layered van der Waals magnetism as a versatile platform for high-efficiency, programmable, field-free superconducting diodes.
\end{abstract}

\maketitle

Symmetry breaking underlies many central phenomena in condensed-matter physics, from topological phases to unconventional superconductivity~\cite{Chiu2016:RMP}. A prominent recent example is the superconducting diode effect (SDE)~\cite{Hu2007:PRL,Wakatsuki2017:SA,Ando2020:Nature,Baumgartner2022:NN,Wu2022:Nature,Nadeem2023:NRP}, in which the critical
supercurrent becomes direction dependent, enabling dissipationless rectification. The SDE requires the simultaneous breaking of inversion and time-reversal symmetries~\cite{Wakatsuki2017:SA,Tokura2018:NC,Jiang2022:NP,Davydova2022:SA} and has attracted strong
interest both as a probe of symmetry-controlled superconductivity and as a building block for ultralow-power superconducting logic and memory.

Superconducting nonreciprocity has been observed in diverse platforms, including noncentrosymmetric superconductors~\cite{Wakatsuki2017:SA,Ando2020:Nature}, Josephson junctions~\cite{Baumgartner2022:NN,Wu2022:Nature,Jeon2022:NM,Pal2022:NP}, and low-dimensional materials with strong
spin--orbit coupling (SOC)~\cite{Itahashi2020:SA,Bauriedl2022:NC}.
In most cases, however, the SDE
appears only under an applied magnetic field~\cite{Baumgartner2022:NN,Bauriedl2022:NC} and is commonly
attributed to finite-momentum Cooper pairing~\cite{Davydova2022:SA,Pal2022:NP,Yuan2022:PNAS,Daido2022:PRL,Ilic2022:PRL}.
At the same time, nonreciprocal signals can also arise from extrinsic mechanisms such as Meissner screening~\cite{Vodolazov2005:PRB,Hou2023:PRL,Moll2023:NP} or current-induced self-fields~\cite{Vasenko1981:JETP,Krasnov1997:PRB,Golod2022:NC}, complicating the identification of intrinsic diode behavior. Achieving a robust, high-efficiency SDE at zero magnetic field therefore remains a key challenge for both fundamental studies and device applications~\cite{Wu2022:Nature,Jeon2022:NM,Ghosh2024:NM,Narita2022:NN,Xiong2024:NC,Qi2025:NC,Anwar2023:CP,Nagata2025:PRL,Zhang2025:SA}.

Van der Waals heterostructures that combine superconductors with magnetic materials provide an appealing route toward field-free
superconducting diodes. In such systems, time-reversal symmetry can be lifted internally via magnetic proximity. Although several
superconductor--magnet hybrids have demonstrated zero-field SDEs~\cite{Jeon2022:NM,Hou2023:PRL,Narita2022:NN,Xiong2024:NC,Zhang2025:SA}, the reported efficiencies are typically
modest, and the underlying physics often relies on relatively conventional magnetic configurations. Realizing large and tunable field-free nonreciprocity calls for coupling superconductivity to more complex magnetic textures.

\begin{figure*}
	\centering
	\vspace{0.2cm}
	\includegraphics*[width=0.76\textwidth]{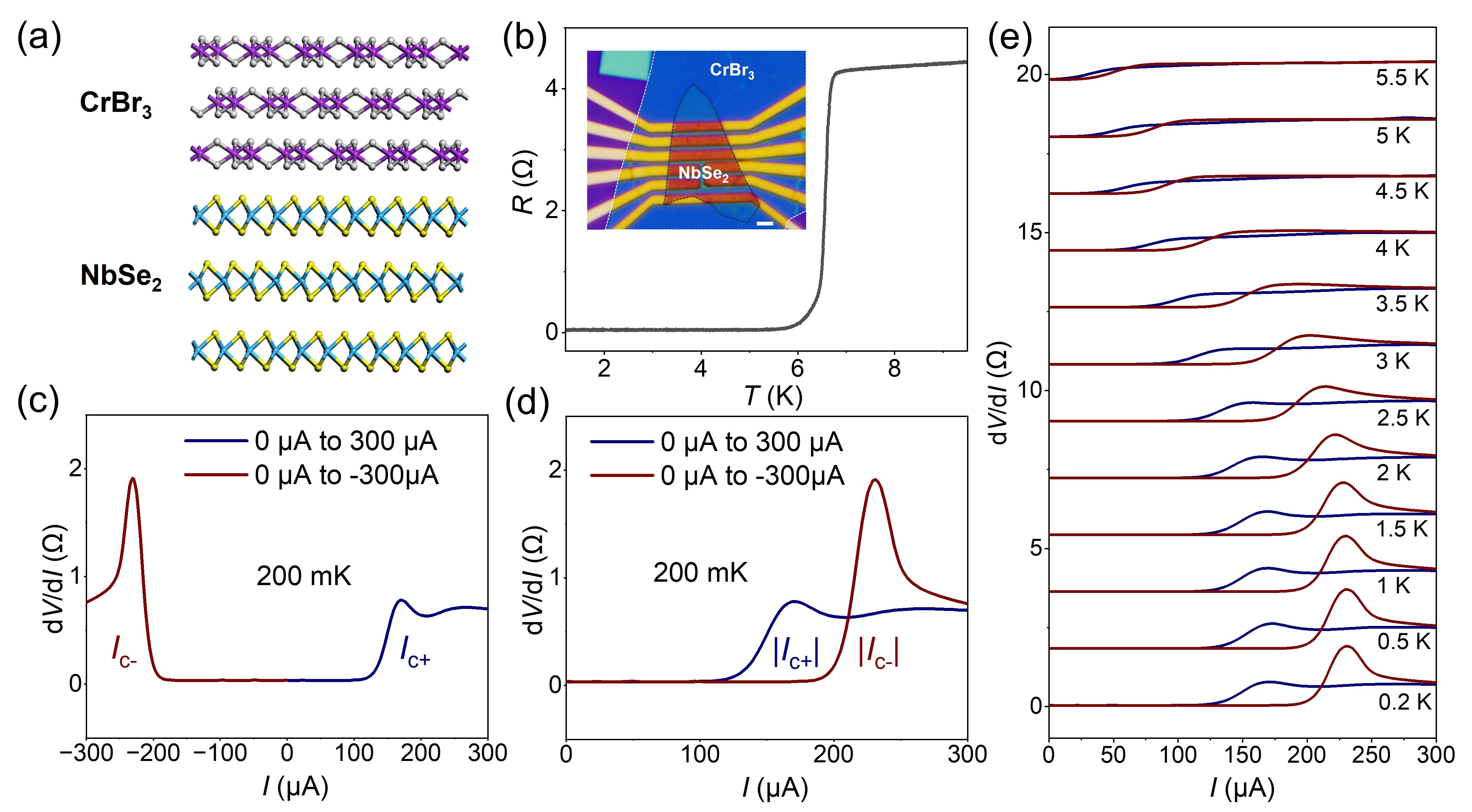}
	\vspace{-0.2cm}
	\caption{Device configuration and superconducting behavior of NbSe$_2$/CrBr$_3$ heterostructure (device~\#1). (a)~Schematic of the crystal structure of the heterostructure. (b)~The resistance--temperature curve of the heterostructure, showing typical superconducting behavior. The resistance is measured from the conducting NbSe$_2$ layer proximitized by the insulating CrBr$_3$. The inset is the optical image of the device. The scale bar is $2~\text{\textmu}\mathrm{m}$. (c)~The differential resistance curves of the heterostructure, highlighting a pronounced superconducting diode effect, where $|I_{\text{c}+}| \neq |I_{\text{c}-}|$. The forward sweep (blue) refers to current swept from $0\,\text{\textmu}\mathrm{A}$ to $+300\,\text{\textmu}\mathrm{A}$ and the backward sweep (red) refers to current swept from $0\,\text{\textmu}\mathrm{A}$ to $-300\,\text{\textmu}\mathrm{A}$. (d)~Replot of the differential resistance curves against the absolute current magnitude ($|I_{\text{c}}|$). (e)~Temperature dependence of the differential resistance curves, plotted in the same mirrored format as (d) and vertically offset for clarity.}
	\label{fig1}
\end{figure*}

Here we address this challenge using a van der Waals (vdW) heterostructure combining the Ising superconductor NbSe$_2$ \cite{Frindt1972:PRL,Xi2016:NP,Wan2023:Nature} with the layered magnetic insulator CrBr$_3$ \cite{Ho1969:PRL,Chen2019:Science,Kim2019:NE,Sun2021:NC,Yang2023:JACS,Yao2023:NC}. Multilayer CrBr$_3$ hosts coexisting ferromagnetic and antiferromagnetic interlayer couplings, giving rise to ferrimagnetic (FIM) states and multistep magnetization reversal \cite{Chen2019:Science,Yang2023:JACS,Yao2023:NC}. When coupled to NbSe$_2$, this layered magnetism generates a history-dependent interfacial exchange field. We observe a large field-free SDE with a zero-field efficiency reaching 35.7\% and a polarity programmable by magnetic-field history. Moreover,upon sweeping a perpendicular magnetic field, the diode polarity undergoes nonmonotonic and hysteretic reversals. Unlike conventional finite-momentum-pairing scenarios, the SDE originates from a symmetry-allowed cubic term odd in the Cooper-pair momentum $q$. Within a generalized Ginzburg--Landau (GL) framework, this term arises from the combined effects of Ising SOC, trigonal warping, and the effective out-of-plane field $B_{\mathrm{eff}}$, producing an even-in-$q$ supercurrent component and hence unequal critical currents while the equilibrium condensate remains at $q=0$. First-principles calculations, magnetic characterization, and micromagnetic simulations further reveal that $B_{\mathrm{eff}}$ is governed by the interfacial magnetization of layered FIM CrBr$_3$, which can be opposite to the net magnetization and evolve nontrivially with magnetic-field history. This mechanism accounts for the observed nonmonotonic and hysteretic polarity reversals and establishes layered vdW magnetism as a versatile route to programmable, field-free superconducting diodes.

Due to the 2D exfoliated nature of the NbSe$_2$/CrBr$_3$ heterostructure, it is fabricated using a standard dry-transfer technique in a nitrogen-filled glovebox ($< 0.01$ ppm O$_{2}$ and H$_{2}$O concentrations). Figures 1(a) and 1(b) illustrate
the schematic and optical image of the heterostructure device. The bottom electrodes are exclusively in contact with the
NbSe$_2$ layer, and the resistance is measured using a standard four-probe method. The resistance-temperature curve is shown in
Fig. 1(b), showing a typical superconducting behavior. Figure 1(c) shows the differential resistance curve of the NbSe$_2$ layer
proximitized by the magnetic insulator CrBr$_3$. The peaks in the differential resistance correspond to the critical current along
the horizontal axis. Here, the critical currents corresponding to the forward and backward current sweeps are defined as \(\lvert I_{\text{c}+}\rvert\) and \(\lvert I_{\text{c}-}\rvert\), respectively. The backward branch (red) from Fig. 1(c) is mirrored into the first quadrant to visually emphasize the difference between the critical currents $|I_{\text{c}+}|$ and $|I_{\text{c}-}|$. The significant asymmetry between
\(\lvert I_{\text{c}+}\rvert\) and \(\lvert I_{\text{c}-}\rvert\) without any magnetic field in Fig. 1(d)
confirms the field-free superconducting diode effect, where the critical current in one direction exceeds that in the opposite direction. Figure 1(e) further demonstrates that the SDE persists throughout the whole superconducting regime. 

The field-free SDE in the heterostructure can be further modulated by
the magnetic field history. Fig. 2(a) shows a map of the differential
resistance as a function of the magnetic field and bias currents during
the forward and backward magnetic field sweep. When the magnetic field
is swept from positive values to zero, the SDE exhibits a positive
polarity, while sweeping the magnetic field from negative values to zero
results in a negative polarity. The corresponding critical currents
\(\lvert I_{\text{c}+}\rvert\) and \(\lvert I_{\text{c}-}\rvert\) are extracted in Fig. 2(b). The
nonreciprocal component of the critical currents
\(\Delta I_{\text{c}} = |I_{\text{c}+}| - |I_{\text{c}-}|\) and corresponding SDE
efficiency
\(\eta = (|I_{\text{c}+}| - |I_{\text{c}-}|)/( |I_{\text{c}+}| + |I_{\text{c}-}| )\) are
plotted in Fig. 2(c).

\begin{figure*}[t]
	\vspace{0.2cm}
	\includegraphics*[width=0.76\textwidth]{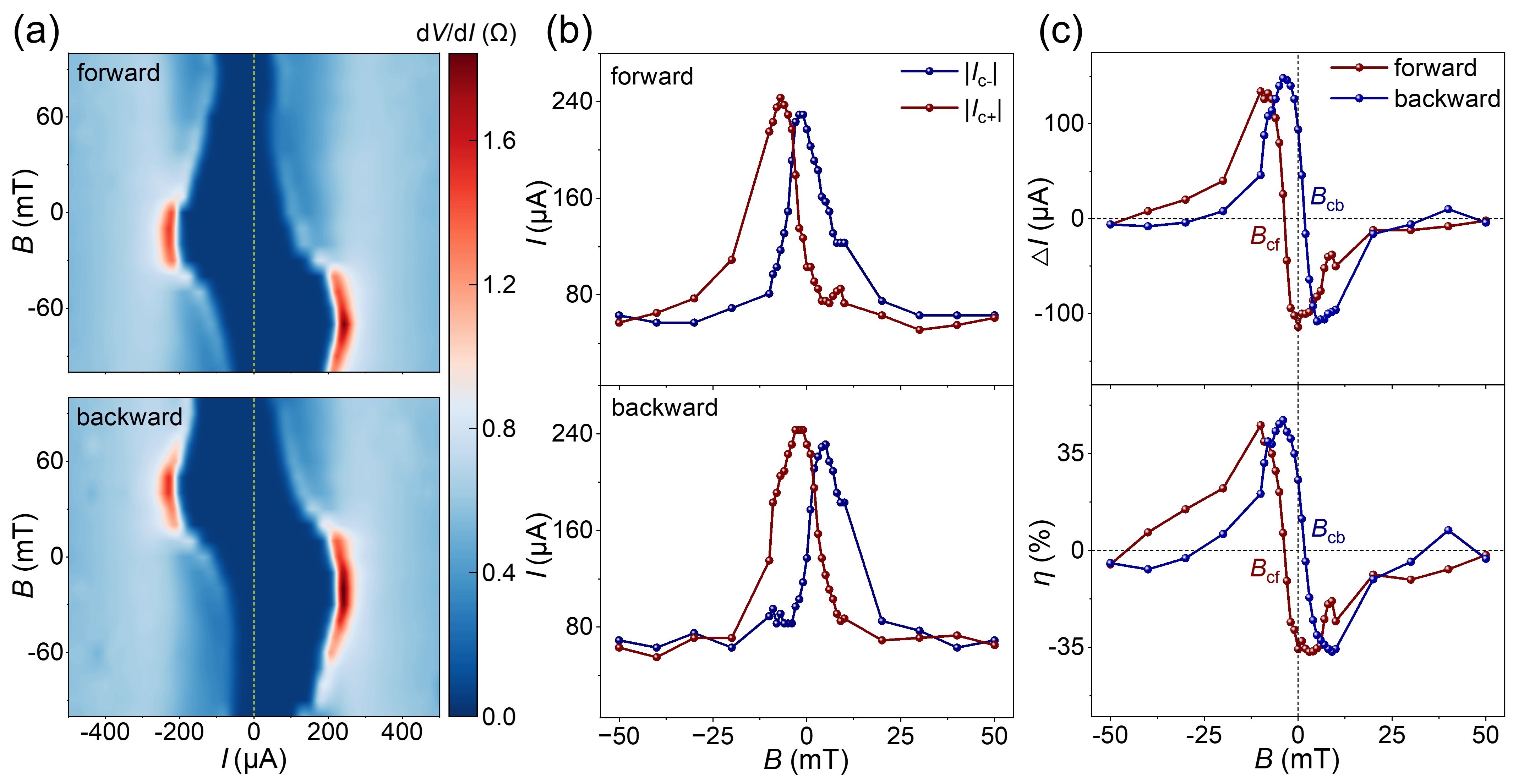}
	\vspace{-0.2cm}
	\caption{The superconducting diode effect tuned by the magnetic field at 200~mK (device~\#1). (a)~Map of the differential resistance as a function of magnetic field and bias current, showing the forward (upper panel) and backward (lower panel) sweep directions of the magnetic field. (b) The magnetic field dependence of the critical currents $|I_{\text{c}+}|$ and $|I_{\text{c}-}|$. (c) The magnetic field dependence of nonreciprocal component of the critical current $\Delta I_{\text{c}}$ (upper panel) and superconducting diode efficiency $\eta$ (lower panel),
		 calculated using $\eta=\bigl(|I_{\text{c}+}|-|I_{\text{c}-}|\bigr)/( |I_{\text{c}+}|+|I_{\text{c}-}| )$. Here, $B_{\text{cb}}$ denotes the value of the external field at which the sign change of the SDE occurs in the backward sweep, and $B_{\text{cf}}$ corresponds to the sign change in the forward sweep.}
	\label{fig2}
\end{figure*}

Figure 2(c) provides key insights into the SDE observed in the NbSe$_2$/CrBr$_3$ heterostructure. First, as previously mentioned, the field-free SDE changes sign depending on the magnetic-field history, i.e., whether the magnetic field is swept forward or backward. Second, after undergoing the magnetic field training, the field-free SDE efficiency reaches $-$35.7\%. In contrast, the original efficiency after zero-field cooling is $-$15.2\%, as shown in Fig. 1(d). Third, the SDE displays distinct behavior under varying magnetic fields. Taking the forward magnetic field sweep as an example, as the field sweeps from $-$50 mT to 0 mT, the SDE efficiency first reaches a maximum positive value of 45.3\% at around $-$10 mT, then rapidly undergoes a sign change, ultimately reaching $-$35.7\% at 0 mT, showing nonmonotonic polarity reversals. Beyond 0 mT, as the field increases to 50 mT, the SDE efficiency gradually decreases. During the reverse sweep from 50 mT to $-$50 mT, the behavior mirrors that of the forward sweep, demonstrating a hysteretic SDE response, as illustrated in Fig. 2(c). \(B_{\text{cb}}\) and \(B_{\text{cf}}\) denote the value of \(B\)
at which the sign change of SDE occurs in backward and forward sweeps of the magnetic field, respectively. It is found that \(B_{\text{cb}}\) \textgreater{} 0 and \(B_{\text{cf}}\) \textless{} 0. Similar behavior is also observed in another device \#2, as shown in Supplementary Material (SM)~\cite{SM}.


To further present the tunability of the zero-field SDE in NbSe$_2$/CrBr$_3$ heterostructure, we performed systematic measurements to characterize its rectification response under different initial magnetization configurations, as illustrated in Fig. 3(a). Application of a 6 T perpendicular magnetic field for initial magnetization induced positive magnetization in the CrBr$_3$ layer, establishing a state wherein the critical current \(\lvert I_{\text{c}+}\rvert > \lvert I_{\text{c}-}\rvert\). With a $705\,\text{\textmu}\mathrm{A}$
excitation current between \(\lvert I_{\text{c}+}\rvert\) and \(\lvert I_{\text{c}-}\rvert\), the device sustained superconductivity across the positive current half-cycle, showing zero voltage, while developing a finite voltage in the negative half-cycle. Conversely, initial magnetization with a $-$6 T field reoriented the CrBr$_3$ magnetization to a negative state, This inverts the
critical current asymmetry to \(\lvert I_{\text{c}-}\rvert > \lvert I_{\text{c}+}\rvert\), thereby resulting in the opposite rectification polarity. These observations of zero-field rectification not only validate diode-like operational characteristics but also establish a strong correlation between the sign of SDE and the magnetization direction of the CrBr$_3$ layer within the heterostructure.

The rectification enabled by the SDE is essential for the realization of dissipationless electronic circuits. Beyond rectification stability, high diode efficiency together with a large $\Delta I_{\text{c}}$ is crucial for practical superconducting electronics. As summarized in Fig. 3(b), which compiles previous reports of field-free SDE~\cite{Wu2022:Nature,Jeon2022:NM,Hou2023:PRL,Ghosh2024:NM,Narita2022:NN,Xiong2024:NC,Qi2025:NC,Anwar2023:CP,Nagata2025:PRL,Zhang2025:SA,Zhao2023:Science}, the NbSe$_2$/CrBr$_3$ heterostructure studied here simultaneously achieves one of the highest zero-field diode efficiencies (35.7\%) and a large $\Delta I_{\text{c}}$. It is worth emphasizing that, unlike actively driven configurations that require continuous external microwave irradiation to dynamically induce nonreciprocity~\cite{Borgongino2025:NL,Wang2026:NP}, the large field-free SDE in our system is stemming from a static magnetic proximity effect without external fields. This combination of rectification performance and atomically thin, van der Waals materials highlights the potential of this system for scalable superconducting electronic applications.

From a symmetry perspective, a field-free SDE requires the simultaneous breaking of time-reversal and inversion symmetries~\cite{Wakatsuki2017:SA,Tokura2018:NC,Jiang2022:NP,Davydova2022:SA}. In the present NbSe$_2$/CrBr$_3$ heterostructure, the inequivalent environments on the two sides of NbSe$_2$ break inversion symmetry, while magnetic proximity to CrBr$_3$ produces a spontaneous out-of-plane exchange field $B_{\text{prox}}$ even at zero applied field~\cite{Kezilebieke2020:Nature}, thereby breaking time-reversal symmetry. This naturally accounts for the emergence of a field-free SDE and explains why the diode polarity depends on magnetic-field sweep history.


Within the conventional ferromagnet/Ising-superconductor picture~\cite{Xiong2024:NC}, $B_{\text{prox}}$ is expected to align with the external magnetic field at the beginning of a sweep due to magnetic coercivity. As a result, the diode polarity should reverse when $B$ cancels $B_{\text{prox}}$, yielding sign-change fields $B_{\text{cf}} > 0$ for forward sweeps and $B_{\text{cb}} < 0$ for backward sweeps. In striking contrast, our NbSe$_2$/CrBr$_3$ devices consistently exhibit the opposite condition, $B_{\text{cf}} < 0$ and $B_{\text{cb}} > 0$, even after exhaustively checking all possible sign conventions for $B$ and $\Delta I_{\text{c}}$ as seen in SM~\cite{SM}. This inverted sign-change behavior cannot be reconciled with a simple ferromagnetic proximity model.

\begin{figure}[t!]
	\centering
	\vspace{0.2cm}
	\includegraphics*[width=0.48\textwidth]{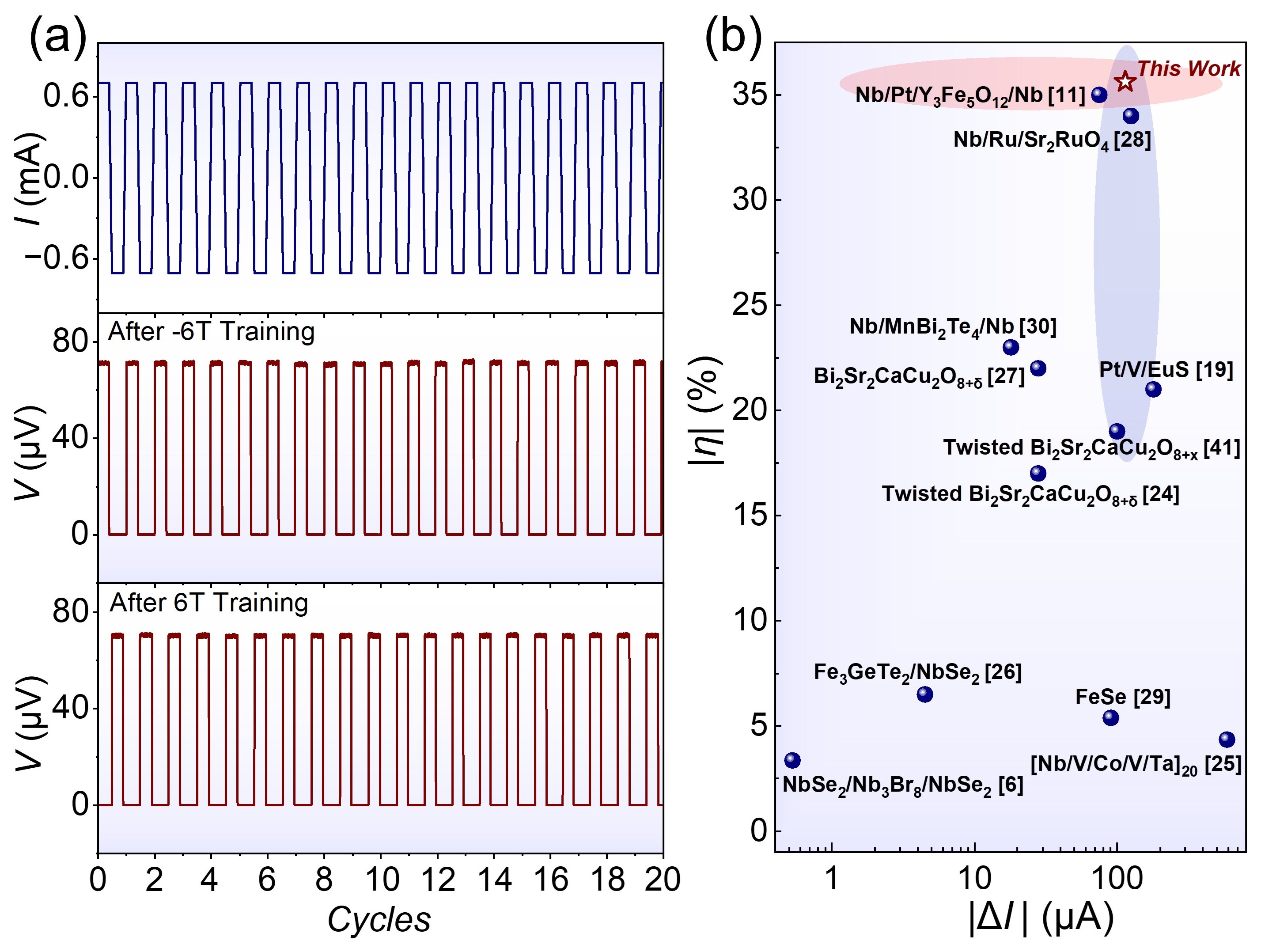}
	\vspace{-0.5cm}
	\caption{(a)~Zero-field rectification of NbSe$_2$/CrBr$_3$ heterostructure after applying perpendicular magnetic fields of $6$~T and $-6$~T in device~\#3. (b)~Comparison of field-free superconducting diode effect $\eta$ and rectification current $\Delta I_{\text{c}}$ from previous reports and this work (device~\#1). This work reaches a field-free diode efficiency of 35.7\%, among the highest reported to date.} 
	\label{fig3}
\end{figure}

While previous theories of intrinsic SDE predict magnetic-field-induced sign reversals of diode efficiency~\cite{Daido2022:PRL,Ilic2022:PRL}, these treatments focus on Rashba-type superconductors and are not directly applicable to Ising superconductors coupled to layered magnets. Other proposals, such as oscillatory SDE behavior driven by valley polarization in twisted graphene~\cite{Hu2023:PRL}, are also inconsistent with our data, as such oscillations are not observed. These discrepancies indicate that a new microscopic mechanism is required to explain the observed behavior. Other trivial mechanisms for the zero-field SDE (e.g., self-fields, Meissner screening, and geometric asymmetry) are ruled out, as shown in SM~\cite{SM}. We therefore first examine the magnetic origin and field history of $B_{\text{prox}}$.

\begin{figure}[t!]
	\centering
	\vspace{0.2cm}
	\includegraphics*[width=0.48\textwidth]{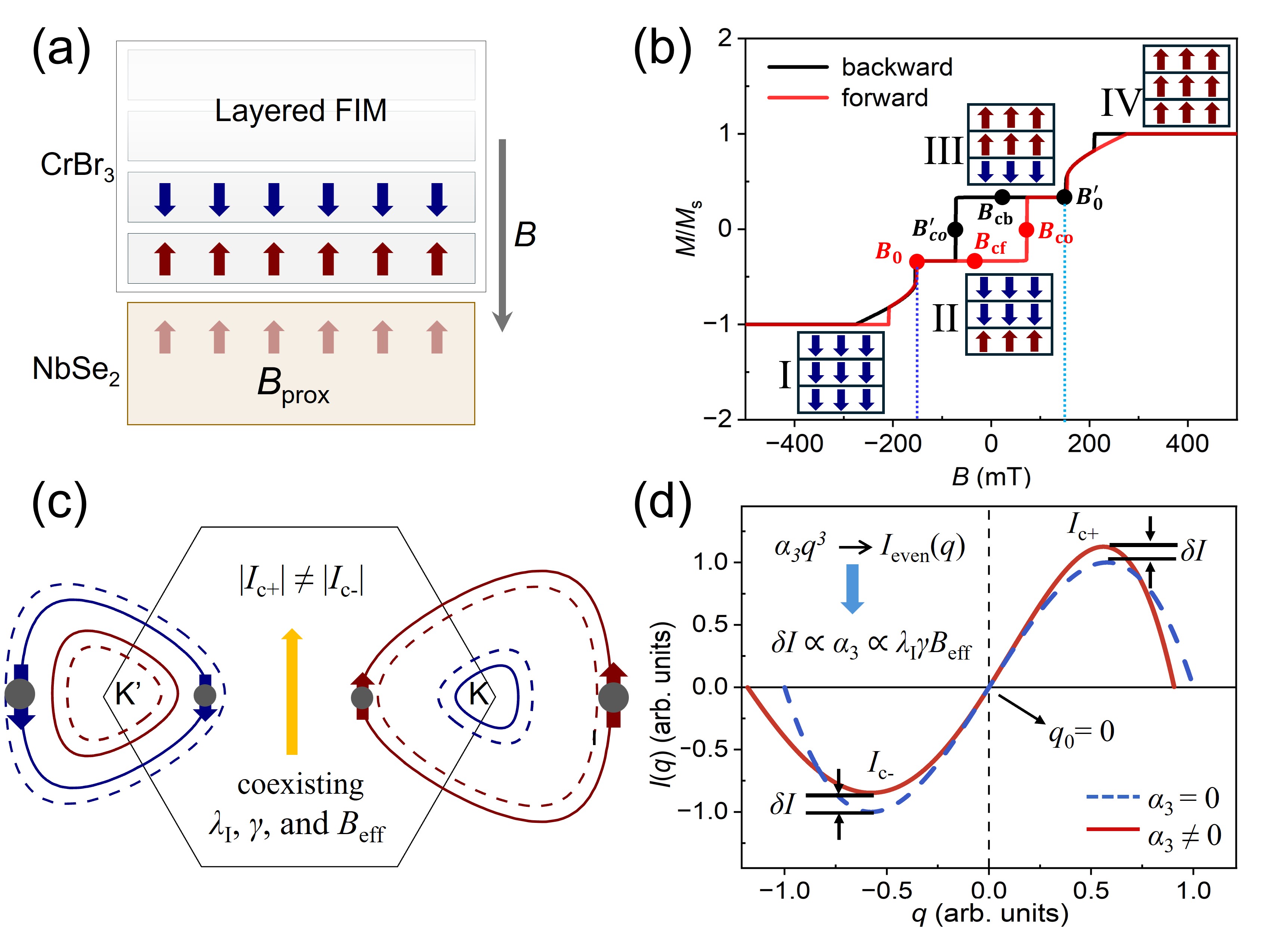}
	\vspace{-0.1cm}
\caption{(a)~Schematic of the proximity-induced exchange field $B_{\mathrm{prox}}$ in NbSe$_2$ from layered ferrimagnetic CrBr$_3$. Arrows denote the magnetization of individual CrBr$_3$ layers. (b)~Micromagnetic simulation of field-driven magnetization reversal in a CrBr$_3$ trilayer. The hysteresis loop exhibits four plateaus (I--IV), reflecting multistep reversal. $B_0$ ($B'_0$) and $B_{\mathrm{co}}$ ($B'_{\mathrm{co}}$) indicate the switching and coercive magnetic fileds that delimit the FIM configurations, while $B_{\mathrm{cf}}$ and $B_{\mathrm{cb}}$ denote the SDE polarity-reversal fields where $B_{\mathrm{eff}}=0$. (c)~Spin-resolved Fermi surfaces of NbSe$_2$ with finite Ising spin--orbit coupling $\lambda_I$, trigonal warping $\gamma$, and effective field $B_{\mathrm{eff}}$. Red (blue) contours denote spin-up (spin-down) states; solid (dashed) contours indicate cases with (without) $B_{\mathrm{cb}}$. (d)~Supercurrent $I(q)$ for finite (solid) and vanishing (dashed) cubic coefficient $\alpha_3$. For $\alpha_3=0$, $I(q)$ is odd in $q$ and $|I_c^+|=|I_c^-|$; finite $\alpha_3$ introduces an even-in-$q$ component, yielding $|I_c^+|-|I_c^-|=2\delta I$.}
\label{fig4}
\end{figure}

In contrast to conventional ferromagnetic proximity scenarios, $B_{\text{prox}}$ in NbSe$_2$/CrBr$_3$ does not simply follow the net magnetization of CrBr$_3$. Our polar RMCD measurements~\cite{SM} demonstrate that multilayer CrBr$_3$ hosts a layered FIM state [Fig.~4(a)], arising from the coexistence of ferromagnetically and antiferromagnetically coupled layers, which is confirmed by our micromagnetic simulations as seen in SM~\cite{SM}. Because magnetic proximity is dominated by the CrBr$_3$ layer adjacent to NbSe$_2$, as further supported by our first-principles calculations~\cite{SM}, the resulting $B_{\text{prox}}$ can acquire a sign opposite to that of the total magnetization of the CrBr$_3$ stack.

Our micromagnetic simulations (Fig.~4(b)) of a CrBr$_3$ trilayer show that a minimal configuration with ferromagnetically aligned upper layers and antiferromagnetic coupling to the bottom layer naturally produces multistep magnetization plateaus. In this configuration, the bottom CrBr$_3$ layer, which is directly interfaced with NbSe$_2$, can remain antiparallel to the $B$ over a finite field window, even when the overall magnetization is only partially reversed. The micromagnetic simulations reproduce the RMCD hysteresis and identify the layered ferrimagnetic configurations of CrBr$_3$, namely the magnetic switching fields $B_0$ ($B_0'$) and $B_{\mathrm{co}}$ ($B'_{\mathrm{co}}$) as shown in Fig.~4(b). The transport-defined sign-reversal fields $B_{\text{cf}}$ and $B_{\text{cb}}$ correspond to compensation points of the low-field superconducting nonreciprocity coefficient, where the interfacial exchange-induced contribution and the external-field induced contribution cancel. Thus, $B_{\text{cf}}$ and $B_{\text{cb}}$ are expected to occur within the ferrimagnetic plateau, that is, between $B_0$ ($B_0'$) and $B_{\mathrm{co}}$ ($B'_{\mathrm{co}}$). During a forward field sweep, $B_{\text{prox}}$ can oppose the applied field $B$, causing $B_{\text{eff}}$ to cross zero at a negative field (\(B_{\text{cf}} < 0\)). During the reverse sweep, the same mechanism shifts the zero crossing to a positive field (\(B_{\text{cb}} > 0\)).

To understand how the magnetic proximity effect generates a nonreciprocal supercurrent, we consider the superconducting free energy of NbSe$_2$ within a generalized Ginzburg--Landau (GL) framework near $T_c$. The CrBr$_3$ interface breaks inversion symmetry, while the effective out-of-plane field $B_{\mathrm{eff}}=B+B_{\mathrm{prox}}$ breaks time-reversal symmetry. Together with the intrinsic Ising spin--orbit coupling and trigonal warping of NbSe$_2$, these ingredients allow the superconducting response to become asymmetric with respect to the Cooper-pair momentum $q$. For a fixed in-plane transport direction, the GL free-energy density can be expanded as (see more details in SM~\cite{SM})
\begin{equation}
 f(q,\Delta)=\left(\alpha_0+\alpha_2q^2+\alpha_3q^3+\cdots\right)|\Delta|^2+\beta|\Delta|^4,
 \label{eq:GL}
\end{equation}
where $q$ is the signed Cooper-pair momentum along the current direction, and $\Delta$ is the superconducting order parameter. For NbSe$_2$ under an out-of-plane effective field, the threefold rotational symmetry forbids the conventional linear-in-$q$ Lifshitz term, while allowing a cubic odd-in-$q$ contribution associated with the trigonal harmonic $q^3\cos 3\theta$. To leading order, its coefficient takes the form
\begin{equation}
 \alpha_3\propto \lambda_I\gamma B_{\mathrm{eff}}\cos 3\theta,
 \label{eq:alpha3}
\end{equation}
where $\lambda_I$ is the Ising spin--orbit coupling, $\gamma$ characterizes trigonal warping, and $\theta$ denotes the transport direction relative to a crystalline axis. Thus, the cubic nonreciprocal term arises from the combined action of Ising spin--orbit coupling, trigonal warping, and the effective magnetic field, as schematically illustrated in Fig.~4(c).

A key consequence of Eq.~\eqref{eq:GL} is that the cubic term generates nonreciprocity without shifting the equilibrium condensate to finite momentum. Because the linear-in-$q$ term is absent, $q=0$ remains a stable equilibrium state. The $q^3$ term instead makes the finite-$q$ current-carrying states asymmetric about equilibrium. Minimizing the free energy with respect to $|\Delta|$ gives $f_{\min}(q)=-\alpha^2(q)/\beta$, and the supercurrent $I(q)\propto \partial f_{\min}/\partial q$ becomes, to first order in $\alpha_3$,
\begin{align}
 I(q)\propto{}&-\left(2\alpha_0\alpha_2q+2\alpha_2^2q^3\right)-\left(3\alpha_0\alpha_3q^2+5\alpha_2\alpha_3q^4\right) \notag\\
 \equiv{}& I_{\mathrm{odd}}(q)+I_{\mathrm{even}}(q).
 \label{eq:Iq}
\end{align}
Equation~\eqref{eq:Iq} makes the origin of the SDE transparent. For $\alpha_3=0$, only $I_{\mathrm{odd}}(q)$ remains, so that $I(q)=-I(-q)$ and the two critical currents have equal magnitudes, $|I_c^+|=|I_c^-|$. A finite $\alpha_3$, however, introduces the even-in-$q$ component $I_{\mathrm{even}}(q)$, breaking this antisymmetry and yielding $|I_c^+|\neq|I_c^-|$, as illustrated in Fig.~4(d). Thus, unlike the conventional helical mechanism in which a linear Lifshitz term shifts the equilibrium condensate to finite momentum, the SDE here originates directly from a cubic odd-in-$q$ term that renders the current-carrying states asymmetric while the equilibrium condensate remains at $q=0$.

This framework also directly connects the diode polarity to the magnetic state of CrBr$_3$. Because $\alpha_3$ is odd in $B_{\mathrm{eff}}$, reversing $B_{\mathrm{eff}}$ reverses $I_{\mathrm{even}}(q)$ and hence the SDE polarity. The unusual polarity reversals can therefore be understood by examining how $B_{\mathrm{eff}}$ evolves during a magnetic-field sweep. In a conventional ferromagnet/superconductor heterostructure, $B_{\mathrm{prox}}$ is expected to approximately follow the net magnetization, leading to a relatively simple compensation between the proximity-induced and external fields. The situation in NbSe$_2$/CrBr$_3$ is qualitatively different.

We therefore identify a non-helical cubic-momentum SDE in which nonreciprocity arises from asymmetric depairing of the two current-carrying branches rather than from a displacement of the equilibrium condensate. This differs from the common helical setting, where a linear Lifshitz invariant first shifts the equilibrium state to $q_0\neq0$ and additional higher-order terms make the depairing currents nonreciprocal~\cite{Davydova2022:SA,Pal2022:NP,Yuan2022:PNAS,Daido2022:PRL,Ilic2022:PRL}. Here $q_0=0$, but the cubic odd-in-$q$ free-energy term breaks the antisymmetry $I(q)=-I(-q)$. Reversing $B_{\text{eff}}$ reverses this even-in-$q$ current component and therefore the diode polarity. Because $B_{\text{prox}}$ is controlled by the interfacial CrBr$_3$ layer, it can oppose the applied field within the layered ferrimagnetic plateaus, naturally accounting for the nonmonotonic and hysteretic polarity reversals.

In summary, we demonstrate a large, programmable field-free SDE in NbSe$_2$/CrBr$_3$ heterostructures, with a zero-field efficiency reaching 35.7\%. A generalized GL description identifies a symmetry-allowed cubic odd-in-momentum term that generates an even-in-$q$ supercurrent component and unequal critical currents while preserving a zero-momentum equilibrium condensate. The nonmonotonic and hysteretic polarity reversals arise from layered ferrimagnetism in CrBr$_3$, for which the interfacial proximity field can evolve differently from, and even oppose, the net magnetization.

Beyond field-free superconducting rectification, our results reveal that the interfacial exchange field in a layered magnet need not be determined by its total magnetization. This principle provides a route to exploring proximity effects in more complex magnetic systems, including recently identified Stoner--Wohlfarth antiferromagnets~\cite{Wang2026:Nature}. More broadly, it may enable unconventional interfacial phenomena in heterostructures combining layered magnets with topological insulators and superconductors~\cite{zhu2026altermagnetic,Zutic2019:MT,chen2026altermagnets,huang2025towards,Zhou2023:NM,Amundsen2024:RMP}, opening opportunities to integrate nonreciprocity, magnetic memory, and topology within a common material platform.

\section{acknowledgments}
\begin{acknowledgments}
B.-C.L. thanks Junyi Zhang, Kam Tuen Law, Xilin Feng, Akito Daido, Zixian Yang for valuable discussions. This work was supported by the National Key Research and Development Program of China (No. 2022YFA1403700), National Natural Science Foundation of China (12474155, 12447163, and 12504058), Guangdong Basic and Applied Basic Research Foundation (Grant No. 2022B1515130005), the Zhejiang Provincial Natural Science Foundation of China (LR25A040001), the China Postdoctoral Science Foundation (2025M773440), and the Shenzhen International Quantum Academy (Grant No. SIQA2024KFKT03, SIQA2025KFKT05).
\end{acknowledgments}

\bibliography{main_Refs}

\end{document}